# Controlled polarization rotation of an optical field in multi-Zeeman-sublevel atoms

Shujing Li[1], Bo Wang[1], Xudong Yang, Yanxu Han, Hai Wang[1*], Min Xiao[1,2] and K.C. Peng[1]

[1] *The State Key Laboratory of Quantum Optics and Quantum Optics Devices, Institute of Opto-Electronics, Shanxi University, Taiyuan, 030006, P.R.China*

[2]*Department of Physics, University of Arkansas, Fayetteville, Arkansas 72701, USA*

ABSTRACT

We investigate, both theoretically and experimentally, the phenomenon of polarization rotation of a weak, linearly-polarized optical (probe) field in an atomic system with multiple three-level electromagnetically induced transparency (EIT) sub-systems. The polarization rotation angle can be controlled by a circularly-polarized coupling beam, which breaks the symmetry in number of EIT subsystems seen by the left- and right-circularly-polarized components of the weak probe beam. A large polarization rotation angle (up to 45 degrees) has been achieved with a coupling beam power of only 15 mW. Detailed theoretical analyses including different transition probabilities in different transitions and Doppler-broadening are presented and the results are in good agreements with the experimentally measured results.

# I. Introduction

The polarization rotation of an optical field or chirality can be caused by the intrinsic helicity of the molecules in the medium or introduced by external electric, magnetic, and optical fields. By introducing asymmetry in the index of refraction for the left- and right-circularly-polarized components of a linearly-polarized optical beam when propagating through the medium, the original linear polarization direction will be rotated. Many schemes have been demonstrated in inducing such chirality in various atomic and molecular systems. The most studied phenomenon in such induced polarization rotation is the magneto-optical effect. External magnetic field can induce linear or nonlinear magneto-optical effects by introducing frequency shifts among various Zeeman sublevels in atomic vapors [1,2], which have let to the development of sensitive magnetometry and nonlinear magneto-optical tomography. Optical fields can also be used to introduce asymmetries in different energy levels, such to change the indices of refraction for the left- and right-circularly-polarized optical components of the probe field. Induced polarization rotations by optical pumping of ground-state Zeeman sublevels with a nonresonant light field [3] and by resonant two-photon dispersion in a three-level cascade atomic system [4] were experimentally demonstrated more than thirty years ago. In recent years after demonstrations of the phenomenon of electromagnetically induced transparency (EIT) (especially with low power diode lasers) [5-7], there were renewed interests and new schemes to achieve polarization rotation of an optical beam controlled by another stronger (coupling or controlling) laser beam based on atomic coherence in multi-level EIT systems. Optical birefringence for a linearly-polarized probe beam was experimentally demonstrated in a three-level cascade EIT system by making use of atomic coherence with a cw, circularly-polarized coupling beam [8], which was later improved to have a lower absorption loss and larger achievable polarization rotation angle at a relatively lower coupling power [9]. Similar schemes to achieve polarization rotation were also reported recently in different atomic systems [10,11]. In these schemes, the asymmetry is introduced by connecting one circularly-polarized component (say $\sigma^+$) of the linearly-polarized probe beam to the circularly-polarized coupling beam ($\sigma^-$) through one degenerate middle level (m=+1), which forms a cascade EIT system with less absorption, and leaving another circularly-polarized probe component (σ-) to be highly absorbed (not connected to the coupling beam). Such schemes suffer from strong circular dichroism and therefore still require high coupling beam power (in the order of $10^4$ W/cm$^2$) [9] to achieve a large polarization rotation angle. A more detailed theoretical study was recently presented to reduce optical absorption in inducing polarization rotation of the above system [12]. Also, there were several schemes proposed to control and enhance magneto-optical polarization rotation of a laser beam by employing another laser beam [13,14]. One of such effects, electromagnetically induced magnetochiral anisotropy in a resonant medium, was recently demonstrated experimentally [15].

A new system to achieve optical polarization rotation of a linearly-polarized weak (probe) light beam controlled by a strong, circularly-polarized coupling laser beam was reported recently [16]. The polarization rotation is mainly induced by the asymmetry in the number of Λ-type EIT sub-systems seen by the left- and right-circularly-polarized components of the probe beam with the circularly-polarized coupling beam. In this paper, we present detailed theoretical calculations with careful considerations for contributions of different Clebsch-Gordan (C-G) coefficients for all multi-Z involved atomic transitions, Stark shifts, and Doppler-broadening due to atomic motion in the vapor cell. We show that the essential contribution for the polarization rotation comes from the asymmetry in the number of EIT sub-systems in such eeman-sublevel atomic system, with secondary contribution from different C-G coefficients for various atomic transitions between different Zeeman sublevels. Such detail studies are necessary to fully understand the exact mechanisms of such observed polarization rotation, which can provide an effective way to optimize and control the polarization rotation angle with low coupling beam power. The mechanism for such achieved large polarization rotation angle with relatively low coupling beam power in this new system is very different from previously studied systems in ladder configuration [8-12]. All-optically-controlled polarization rotation can have important applications in dynamic wave-plates, such as half-wave plate, for optical communication and quantum information processing. More detailed experimental studies are also presented and compared with the theoretical calculations.

The rest of the paper is organized as following. Section II presents the theoretical calculations including different C-G coefficients for different transitions and Doppler-broadening, and calculates the degree of polarization rotation as functions of various experimental parameters. Section III describes the experimental setup, the method of detecting the polarization rotation angle, and the experimental procedure. Section IV compares the experimentally measured data with the theoretically calculated results, and provides some discussions. A more detailed discussion on various contributions to the measured polarization rotation angle from different mechanisms is given in Section V. Section VI serves as a conclusion.

## II. Theoretical model

We present a theoretical model to calculate the polarization rotation angle of a linearly-polarized probe field propagating in a multi-Zeeman-sublevel atomic medium by using density-matrix equations. The theoretical calculations take into account different C-G coefficients for different transitions, ac Stark shifts, and Doppler-broadening. The degree of polarization rotation as functions of various experimental parameters is also calculated. The relevant atomic levels of $^{87}$Rb atom in D1 line are shown in Fig. 1. We denote the Zeeman sub-levels of $5S_{1/2}$, F=1 as $|a_i>$ (i=1,2,3 for m=-1, 0, +1), of $5S_{1/2}$, F=2 as $|b_j>$ (j=1-5 for m=-2, -1, 0, +1, +2), and of $5P_{1/2}$, F'=2 as $|c_k>$ (k=1-5 for m=-2, -1, 0, +1, +2), respectively. The basic multi-level atomic system consists of 13 Zeeman sublevels. When both the probe and coupling laser beams are linearly polarized, this system can be considered as the superposition of several simple three-level Λ-type EIT sub-systems [17]. Here, we let the coupling beam (with frequency $\omega_c$) to be a left-circularly-polarized ($\sigma^-$) beam driving the $|b_{j+1}>$ to $|c_j>$ transitions. The probe beam (with frequency $\omega_p$) is a linearly-polarized laser beam consisting of two circularly-polarized components ($\sigma^-$ and $\sigma^+$), which are near resonance with transitions between levels $|a_i>$ and $|c_k>$. The left-circularly-polarized probe beam ($\sigma^-$) couples to the $|a_i>$ to $|c_i>$ transitions and the right-circularly-polarized beam ($\sigma^+$) couples to the $|a_i>$ to $|c_{i+2}>$ transitions. In this case, the system may be considered as five three-level Λ-type EIT sub-systems which consist of levels $|a_1\rangle$-$|c_1\rangle$-$|b_2\rangle$, $|a_2\rangle$-$|c_2\rangle$-$|b_3\rangle$, $|a_3\rangle$-$|c_3\rangle$-$|b_4\rangle$, $|a_1\rangle$-$|c_3\rangle$-$|b_4\rangle$ and $|a_2\rangle$-$|c_4\rangle$-$|b_5\rangle$, respectively. The first three EIT sub-systems are coupled by the left-circularly-polarized coupling beam and the left-circularly-polarized component of the probe beam, while the last two EIT sub-systems are coupled by the left-circularly-polarized coupling beam and the right-circularly-polarized component of the probe beam. The asymmetry in the number of EIT sub-systems is the main reason for generating the polarization rotation of the probe light in this system. For each EIT sub-system (for example, $|a_1\rangle$-$|c_1\rangle$-$|b_2\rangle$), the first-order Doppler effect is eliminated by the two-photon Doppler-free configuration (i.e. co-propagating in such Λ-type system) [6], the atomic coherence effect is significant in each EIT sub-system.

In the interaction picture, and under the dipole and rotating-wave approximations, the Hamiltonian for this system can be described by

$$\hat{H}_{\text{int}} = -\hbar\sum_{i=1}^{5}\Delta\omega_p\,|c_i\rangle\langle c_i| - \hbar\sum_{i=1}^{4}(\Delta\omega_p - \Delta\omega_{ci})\,|b_{i+1}\rangle\langle b_{i+1}| - \frac{\hbar}{2}[\sum_{i=1}^{3}\Omega_{pi}^{-}\,|c_i\rangle\langle a_i|$$
$$+\sum_{i=1}^{3}\Omega_{pi}^{+}\,|c_{i+2}\rangle\langle a_i| + c.c.] - \frac{\hbar}{2}[\sum_{i=1}^{4}\Omega_{ci}^{-}\,|c_i\rangle\langle b_{i+1}| + c.c.]\ , \quad (1)$$

where $\Delta\omega_p = \omega_p - \omega_{ac}$ and $\Delta\omega_{ci} = (\omega_c - \omega_{ci,bi+1})$ are probe and coupling frequency detunings, respectively. $\omega_p$ is the frequency of the probe laser and $\omega_c$ is the frequency of the coupling laser, $\omega_{ac}$ is the transition frequency from atomic levels $|a\rangle$ to $|c\rangle$, while $\omega_{ci,bi+1}$ is the transition frequency from atomic levels $|b_{i+1}\rangle$ to $|c_i\rangle$. $\Omega_{pi}^{-} = -\mu_{ci,ai}E_p^{-}/\hbar$ ($i$=1,2,3), $\Omega_{pi}^{+} = -\mu_{ci+2,ai}E_p^{+}/\hbar$ ($i$=1,2,3), and $\Omega_{ci}^{-} = -\mu_{ci,bi+1}E_c^{-}/\hbar$ ($i$=1,2,3,4) are the Rabi frequencies of the left-circularly-polarized probe beam, right-circularly-polarized probe beam, and left-circularly-polarized coupling beam for various transitions among different Zeeman sub-levels, respectively. The dipole moments $\mu_{i,j}$ are different for different transitions since the Clebsch-Gordan coefficients of the various transitions among different Zeeman sub-levels are different [18], which make the Rabi frequencies different. We have taken into account the differences in Rabi frequencies in following calculations. The evolution of the atomic variables in the system is governed by the master equation

$$\frac{\partial\hat{\rho}}{\partial t} = -\frac{i}{\hbar}\left[\hat{H}_{\text{int}}, \hat{\rho}\right] + \left(\frac{\partial\hat{\rho}}{\partial t}\right)_{inc}, \quad (2)$$

where the first term results from the coherent interactions and the second term represents dampings due to spontaneous and other irreversible processes [19]. This model involves 13 atomic sub-levels, and therefore requires 169 equations including 169 density-matrix elements to fully describe the changes of the atomic variables. One can quickly see that only 75 equations including 75 density-matrix elements are needed in calculating the probe susceptibilities, and the rest 94 equations are trivial. To easily understand and calculate the polarization rotation of the probe field from these 75 equations, it is better to have approximate expressions for the probe susceptibilities. To derive these simplified expressions, we write down the relevant density-matrix equations as following:

$$
\begin{aligned}
\dot{\tilde{\rho}}_{c1,a1} &= (i\Delta\omega_p - \gamma_{ca})\tilde{\rho}_{c1,a1} + i\frac{\Omega_{p1}^-}{2}(\tilde{\rho}_{a1,a1} - \tilde{\rho}_{c1,c1}) + i\frac{\Omega_{c1}^-}{2}\tilde{\rho}_{b2,a1} - i\frac{\Omega_{p1}^+}{2}\tilde{\rho}_{c1,c3} \ , \\
\dot{\tilde{\rho}}_{b2,a1} &= [i(\Delta\omega_p - \Delta\omega_{c1}) - \gamma_{ba}]\tilde{\rho}_{b2,a1} + i\frac{\Omega_{c1}^{-*}}{2}\tilde{\rho}_{c1,a1} - i\frac{\Omega_{p1}^-}{2}\tilde{\rho}_{b2,c1} - i\frac{\Omega_{p1}^+}{2}\tilde{\rho}_{b2,c3} \ , \\
\dot{\tilde{\rho}}_{c2a2} &= (i\Delta\omega_p - \gamma_{ca})\tilde{\rho}_{c2a2} + i\frac{\Omega_{p2}^-}{2}(\tilde{\rho}_{a2a2} - \tilde{\rho}_{c2c2}) + i\frac{\Omega_{c2}^-}{2}\tilde{\rho}_{b3a2} - i\frac{\Omega_{p2}^+}{2}\tilde{\rho}_{c2c4} \ , \\
\dot{\tilde{\rho}}_{b3,a2} &= [i(\Delta\omega_p - \Delta\omega_{c2}) - \gamma_{ba}]\tilde{\rho}_{b3,a2} + i\frac{\Omega_{c2}^{-*}}{2}\tilde{\rho}_{c2,a2} - i\frac{\Omega_{p2}^-}{2}\tilde{\rho}_{b3,c2} - i\frac{\Omega_{p2}^+}{2}\tilde{\rho}_{b3,c4} \ , \\
\dot{\tilde{\rho}}_{c4,a2} &= (i\Delta\omega_p - \gamma_{ca})\tilde{\rho}_{c4,a2} + i\frac{\Omega_{p2}^+}{2}(\tilde{\rho}_{a2,a2} - \tilde{\rho}_{c4,c4}) + i\frac{\Omega_{c4}^-}{2}\tilde{\rho}_{b5,a2} - i\frac{\Omega_{p2}^-}{2}\tilde{\rho}_{c4,c2} \ , \\
\dot{\tilde{\rho}}_{b5,a2} &= [i(\Delta\omega_p - \Delta\omega_{c4}) - \gamma_{ba}]\tilde{\rho}_{b5,a2} + i\frac{\Omega_{c4}^{-*}}{2}\tilde{\rho}_{c4,a2} - i\frac{\Omega_{p2}^-}{2}\tilde{\rho}_{b5,c2} - i\frac{\Omega_{p2}^+}{2}\tilde{\rho}_{b5,c4} \ , \\
\dot{\tilde{\rho}}_{c5,a3} &= (i\Delta\omega_p - \gamma_{ca})\tilde{\rho}_{c5,a3} + i\frac{\Omega_{p3}^+}{2}(\tilde{\rho}_{a3,a3} - \tilde{\rho}_{c5,c5}) - i\frac{\Omega_{p3}^-}{2}\tilde{\rho}_{c5,c3} \ , \qquad (3) \\
\dot{\tilde{\rho}}_{c3,a3} &= (i\Delta\omega_p - \gamma_{ca})\tilde{\rho}_{c3,a3} + i\frac{\Omega_{p3}^-}{2}(\tilde{\rho}_{a3,a3} - \tilde{\rho}_{c3,c3}) + i\frac{\Omega_{c3}^-}{2}\tilde{\rho}_{b4,a3} + i\frac{\Omega_{p1}^+}{2}\tilde{\rho}_{a1,a3} - i\frac{\Omega_{p3}^+}{2}\tilde{\rho}_{c3,c5} \ , \\
\dot{\tilde{\rho}}_{b4,a3} &= [i(\Delta\omega_p - \Delta\omega_{c3}) - \gamma_{ba}]\tilde{\rho}_{b4,a3} + i\frac{\Omega_{c3}^{-*}}{2}\tilde{\rho}_{c3,a3} - i\frac{\Omega_{p3}^-}{2}\tilde{\rho}_{b4,c3} - i\frac{\Omega_{p3}^+}{2}\tilde{\rho}_{b4,c5} \ , \\
\dot{\tilde{\rho}}_{c3,a1} &= (i\Delta\omega_p - \gamma_{ca})\tilde{\rho}_{c3,a1} + i\frac{\Omega_{p1}^+}{2}(\tilde{\rho}_{a1,a1} - \tilde{\rho}_{c3,c3}) + i\frac{\Omega_{p3}^-}{2}\tilde{\rho}_{a3,a1} + i\frac{\Omega_{c3}^-}{2}\tilde{\rho}_{b4,a1} - i\frac{\Omega_{p1}^-}{2}\tilde{\rho}_{c3,c1} \ , \\
\dot{\tilde{\rho}}_{b4,a1} &= [i(\Delta\omega_p - \Delta\omega_{c3}) - \gamma_{ba}]\tilde{\rho}_{b4,a1} + i\frac{\Omega_{c3}^{-*}}{2}\tilde{\rho}_{c3,a1} - i\frac{\Omega_{p1}^-}{2}\tilde{\rho}_{b4,c1} - i\frac{\Omega_{p1}^+}{2}\tilde{\rho}_{b4,c3} \ , \\
\dot{\tilde{\rho}}_{c1,a3} &= [i\Delta\omega_p - \gamma_{ca}]\tilde{\rho}_{c1,a3} + i\frac{\Omega_{p1}^-}{2}\tilde{\rho}_{a1,a3} + i\frac{\Omega_{c1}^-}{2}\tilde{\rho}_{b2,a3} - i\frac{\Omega_{p3}^-}{2}\tilde{\rho}_{c1,c3} - i\frac{\Omega_{p3}^+}{2}\tilde{\rho}_{c1,c5} \ , \\
\dot{\tilde{\rho}}_{a1,c5} &= [-i\Delta\omega_p - \gamma_{ca}]\tilde{\rho}_{a1,c5} + i\frac{\Omega_{p1}^{-*}}{2}\tilde{\rho}_{c1,c5} + i\frac{\Omega_{p1}^{+*}}{2}\tilde{\rho}_{c3,c5} - i\frac{\Omega_{p3}^{+*}}{2}\tilde{\rho}_{a1,a3} \ ,
\end{aligned}
$$

$$\dot{\tilde{\rho}}_{b2,a3} = [i(\Delta\omega_p - \Delta\omega_{c1}) - \gamma_{ba}]\tilde{\rho}_{b2,a3} + i\frac{\Omega_{c1}^{-*}}{2}\tilde{\rho}_{c1,a3} - i\frac{\Omega_{p3}^{-}}{2}\tilde{\rho}_{b2,c3} - i\frac{\Omega_{p3}^{+}}{2}\tilde{\rho}_{b2,c5} \ ,$$

$$\dot{\tilde{\rho}}_{a1,a3} = -i\gamma_a\tilde{\rho}_{a1,a3} + (\frac{\sqrt{6}}{12}\tilde{\rho}_{c1,c3} + \frac{1}{4}\tilde{\rho}_{c2,c4} + \frac{\sqrt{6}}{12}\tilde{\rho}_{c3,c5})\Gamma + i\frac{\Omega_{p1}^{-*}}{2}\tilde{\rho}_{c1,a3} + i\frac{\Omega_{p1}^{+*}}{2}\tilde{\rho}_{c3,a3} - i\frac{\Omega_{p3}^{-}}{2}\tilde{\rho}_{a1,c3} - i\frac{\Omega_{p3}^{+}}{2}\tilde{\rho}_{a1,c5} \ ,$$

where the density-matrix elements $\rho_{\alpha i,\beta j} = \langle \alpha_i | \hat{\rho} | \beta_j \rangle$. *α* and *β* denote the levels *a, b,* or *c* and *i, j* stand for Zeeman sub-level subscripts *1,2,3,4,5*. Decay rates $\gamma_{ij}$ describe decays of populations and coherences. In the absence of outside fields, the Zeeman sub-levels $|b_1\rangle$, $|b_2\rangle$, $|b_3\rangle$, $|b_4\rangle$ and $|b_5\rangle$ are degenerate. When the strong coupling laser beam couples to the transitions from levels $|b_{j+1}\rangle$ to $|c_j\rangle$, it also interacts with the transitions from levels $|b_3\rangle$, $|b_4\rangle$ and $|b_5\rangle$ to $5P_{1/2}$, F'=1 (m=-1, 0, +1) levels with a frequency detuning of $\Delta = 2\pi \times 816 MHz$. Such interactions induce different ac Stark shifts $\hbar\delta_{b3} = \hbar\frac{|\Omega_{C,b3}^{-}|^2}{4\Delta}$, $\hbar\delta_{b4} = \hbar\frac{|\Omega_{C,b4}^{-}|^2}{4\Delta}$, and $\hbar\delta_{b5} = \hbar\frac{|\Omega_{C,b5}^{-}|^2}{4\Delta}$ for the energy levels $|b_3\rangle$, $|b_4\rangle$, and $|b_5\rangle$ [20] (where $\Omega_{C,bi}^{-} = -\mu_{m,bi}E_c^{-}/\hbar$ are the Rabi frequencies of the coupling beam for the transitions from levels $b_i$ (i=3,4,5) to Zeeman sublevels *m*=-1,0,+1 of $5P_{1/2}$, F'=1), respectively. Such ac Stark shifts lift the degeneracy of the Zeeman sublevels $b_i$ (i=3,4,5) and change the atomic transition frequencies from $|b_{i+1}\rangle$ to $|c_i\rangle$ (*i*=2,3,4) to be $\omega_{ci,bi+1} = (\omega_{c,b} + \delta_{bi+1})$ (*i*=2,3,4). $\omega_{c,b}$ is the degenerate transition frequency from atomic levels *b* to *c* when ac Stark shifts are not considered. Therefore, the detunings of the coupling beam from atomic transitions between Zeeman sublevels $|b_{i+1}\rangle$ and $|c_i\rangle$ (*i*=2,3,4) are given by: $\Delta\omega_{ci} = \omega_c - (\omega_{c,b} + \delta_{bi+1}) = \Delta\omega_c - \delta_{bi+1}$. $\Delta\omega_c = \omega_c - \omega_{c,b}$ is the detuning of the coupling beam from atomic transition between levels $|b\rangle$ and $|c\rangle$ in the case of degeneracy for the Zeeman sublevels in $|b\rangle$. Since the probe field is very weak compared to the coupling beam, we can neglect the second order term in $\Omega_p$ and solve Eq. (3) in the steady state [6,17] to obtain the following expressions for the probe beam matrix elements:

$$\tilde{\rho}_{c1,a1} = \frac{i\frac{1}{2}\Omega_{p1}^{-}}{\gamma_{ca} - i\Delta\omega_p + \frac{|\Omega_{c1}^{-}|^2/4}{\gamma_{ba} - i\Delta_c}}\tilde{\rho}_{a1,a1} \ ,$$

$$\tilde{\rho}_{c2,a2} = \frac{i\frac{1}{2}\Omega_{p2}^{-}}{\gamma_{ca} - i\Delta\omega_p + \frac{\left|\Omega_{c2}^{-}\right|^2/4}{\gamma_{ba} - i(\Delta_c + \delta_{b3})}}\tilde{\rho}_{a2,a2} \ ,$$

$$\tilde{\rho}_{c3,a3} = \frac{i\frac{1}{2}\Omega_{p3}^{-}}{\gamma_{ca} - i\Delta\omega_p + \frac{\left|\Omega_{c3}^{-}\right|^2/4}{\gamma_{ba} - i(\Delta_c + \delta_{b4})}}\tilde{\rho}_{a3,a3} \ , \qquad (4)$$

$$\tilde{\rho}_{c3,a1} = \frac{i\frac{1}{2}\Omega_{p1}^{+}}{\gamma_{ca} - i\Delta\omega_p + \frac{\left|\Omega_{c3}^{-}\right|^2/4}{\gamma_{ba} - i(\Delta_c + \delta_{b4})}}\tilde{\rho}_{a1,a1} \ ,$$

$$\tilde{\rho}_{c4,a2} = \frac{i\frac{1}{2}\Omega_{p2}^{+}}{\gamma_{ca} - i\Delta\omega_p + \frac{\left|\Omega_{c4}^{-}\right|^2/4}{\gamma_{ba} - i(\Delta_c + \delta_{b5})}}\tilde{\rho}_{a2,a2} \ ,$$

$$\tilde{\rho}_{c5,a3} = \frac{i\frac{1}{2}\Omega_{p3}^{+}}{\gamma_{ca} - i\Delta\omega_p}\tilde{\rho}_{a3,a3} \ .$$

$\Delta_c = \Delta\omega_p - \Delta\omega_c$ is the two-photon frequency detuning. The values of the ground-state populations ($\rho_{a1a1}$, $\rho_{a2a2}$, $\rho_{a3a3}$) can be calculated by numerically solving the 75 density-matrix equations. When the probe and coupling beam frequency detunings $\Delta\omega_p = \Delta\omega_c = 0$, and the Rabi frequency of the probe beam is 2π×10 MHz, we calculated the ground-state populations to be $\rho_{a1a1}(\infty) = 0.219$, $\rho_{a2a2}(\infty) = 0.228$, $\rho_{a3a3}(\infty) = 0.066$ for the coupling Rabi frequency of $\Omega_c$=2π×60 MHz; $\rho_{a1a1}(\infty) = 0.226$, $\rho_{a2a2}(\infty) = 0.233$, $\rho_{a3a3}(\infty) = 0.066$ for $\Omega_c$=2π×80 MHz; and $\rho_{a1a1}(\infty) = 0.229$, $\rho_{a2a2}(\infty) = 0.235$, $\rho_{a3a3}(\infty) = 0.065$ for $\Omega_c$=2π×100 MHz. The expressions for the susceptibilities of the atomic medium are given by [6]

$$\chi_{ci,aj} = \frac{2N\mu_{ci,aj}}{\varepsilon_0 E_P}\rho_{ci,aj} \ . \qquad (5)$$

To match the theoretical calculations with the experimentally measured results, the Doppler effect due to atomic motion needs to be taken into account by integrating over the atomic velocity distribution [6,17]. If an atom moves against the propagation direction of the probe and coupling beams with velocity $u$, the one-photon frequency detunings will change as:

$$\Delta\omega_p \to \Delta\omega_p + \omega_p \frac{u}{c} , \qquad \Delta\omega_c \to \Delta\omega_c + \omega_c \frac{u}{c} .$$

In our experimental scheme, the coupling and probe beams co-propagate in the atomic cell, which eliminates the first-order Doppler broadening in two-photon frequency detuning ( $\Delta_c = \Delta\omega_p - \Delta\omega_c$ ) [6]. After considering the Doppler effect, the susceptibilities $\chi^-_{c1,a1}$, $\chi^-_{c2,a2}$, $\chi^-_{c3,a3}$ for the left-circularly-polarized component of the probe beam (corresponding to the transitions from levels $|a_1>$ to $|c_1>$, $|a_2>$ to $|c_2>$, and $|a_3>$ to $|c_3>$, respectively) can be written as

$$\chi^-_{c1,a1} = \frac{i}{\hbar\varepsilon_0}\mu^2_{c1,a1}\rho_{a1,a1}F_1 , \qquad (6a)$$

$$\chi^-_{c2,a2} = \frac{i}{\hbar\varepsilon_0}\mu^2_{c2,a2}\rho_{a2,a2}F_2 , \qquad (6b)$$

$$\chi^-_{c3,a3} = \frac{i}{\hbar\varepsilon_0}\mu^2_{c3,a3}\rho_{a3,a3}F_3 . \qquad (6c)$$

The susceptibilities $\chi^+_{c3,a1}$, $\chi^+_{c4,a2}$, $\chi^+_{c5,a3}$ for the right-circularly-polarized component of the probe beam (corresponding to the transitions from levels $|a_1>$ to $|c_3>$, $|a_2>$ to $|c_4>$, and $|a_3>$ to $|c_5>$, respectively) are

$$\chi^+_{c3,a1} = \frac{i}{\hbar\varepsilon_0}\mu^2_{c3,a1}\rho_{a1,a1}F_4 , \qquad (6d)$$

$$\chi^+_{c4,a2} = \frac{i}{\hbar\varepsilon_0}\mu^2_{c4,a2}\rho_{a2,a2}F_5 , \qquad (6e)$$

$$\chi^+_{c5,a3} = \frac{i}{\hbar\varepsilon_0}\mu^2_{c5,a3}\rho_{a3,a3}F_6 . \qquad (6f)$$

The frequency-dependent factors are given by

$$F_1 = \int_{-\infty}^{+\infty} \frac{1}{\gamma_{ca} - i\Delta\omega_p - i\frac{\omega_p}{c}u + \frac{\left|\Omega^-_{c1}\right|^2 / 4}{\gamma_{ba} - i\Delta_c}} N(u)du ,$$

$$F_2 = \int_{-\infty}^{+\infty} \frac{1}{\gamma_{ca} - i\Delta\omega_p - i\frac{\omega_p}{c}u + \frac{\left|\Omega^-_{c2}\right|^2 / 4}{\gamma_{ba} - i(\Delta_c + \delta_{b3})}} N(u)du ,$$

$$F_3 = \int_{-\infty}^{+\infty} \frac{1}{\gamma_{ca} - i\Delta\omega_p - i\frac{\omega_p}{c}u + \frac{\left|\Omega^-_{c3}\right|^2 / 4}{\gamma_{ba} - i(\Delta_c + \delta_{b4})}} N(u)du ,$$

$$F_4 = \int_{-\infty}^{+\infty} \frac{1}{\gamma_{ca} - i\Delta\omega_p - i\frac{\omega_p}{c}u + \frac{\left|\Omega^-_{c3}\right|^2 / 4}{\gamma_{ba} - i(\Delta_c + \delta_{b4})}} N(u)du , \qquad (7)$$

$$F_5 = \int_{-\infty}^{+\infty} \frac{1}{\gamma_{ca} - i\Delta\omega_p - i\frac{\omega_p}{c}u + \frac{\left|\Omega_{c4}^-\right|^2 / 4}{\gamma_{ba} - i(\Delta_c + \delta_{b5})}} N(u)du \quad ,$$

$$F_6 = \int_{-\infty}^{+\infty} \frac{1}{\gamma_{ca} - i\Delta\omega_p - i\frac{\omega_p}{c}u} N(u)du \quad .$$

$N(u)$ is the velocity distribution assumed to obey the Maxwellian distribution as

$$N(u)du = \frac{N_0}{u\sqrt{\pi}} e^{-u^2/V^2} du \quad , \tag{8}$$

where $V/\sqrt{2}$ is the root-mean-square atomic velocity and $N_0$ is the total atomic density of the vapor.

The total susceptibilities for the right- and left-circularly-polarized component of the probe field can be written as

$$\chi_p^- = \chi_{c1,a1}^- + \chi_{c2,a2}^- + \chi_{c3,a3}^- \quad ,$$
$$\chi_p^+ = \chi_{c3,a1}^+ + \chi_{c4,a2}^+ + \chi_{c5,a3}^+ \quad . \tag{9}$$

The polarization rotation angle for the probe beam is defined as $\phi = \frac{\pi}{\lambda}(n_p^+ - n_p^-)d$ [1], where $d$ is the length of the atomic cell and $\lambda$ is the wavelength of the probe beam. $n_p^+$ and $n_p^-$ are the indices of refraction for the right- and left-circularly-polarized probe field components with $n_p^\pm = \sqrt{1 + \mathrm{Re}(\chi_p^\pm)}$. Under the condition that $\chi_p^\pm$ are much smaller than 1, the polarization rotation angle can be approximately written as,

$$\phi \approx \frac{\pi}{2\lambda}\mathrm{Re}(\chi_p^+ - \chi_p^-)d \quad . \tag{10}$$

We numerically calculated the polarization rotation angle as a function of the probe field frequency detuning according to Eq. (10) and found that significant polarization rotation occurs near two-photon EIT resonance. As shown in Fig. 2, there are two maximum polarization rotation angles corresponding to the two dispersion-like peaks. Figure 3 plots the polarization rotation angles at the two dispersion-like peaks as a function of coupling beam Rabi frequency. These results show that the polarization rotation angle ϕ strongly depends on the EIT-resonance shapes for the two circularly-polarized probe field components.

The polarization rotation angle ϕ is induced by the asymmetry for the right- and left-circularly-polarized components of the probe beam i.e. the difference between the two susceptibilities $\chi_p^+$ and $\chi_p^-$. For the present multi-three-level EIT sub-systems, as shown in Fig.1, it is easy to see that the asymmetry for the right- and left-circularly-polarized components of the probe beam includes the asymmetry in number of EIT sub-systems and different transition strengths induced by different transition dipole moments in Zeeman sub-levels (due to different C-G coefficients).

To understand the mechanisms of causing the polarization rotation in the present system, we have carefully analyzed the differences between $\chi_p^+$ and $\chi_p^-$. $\chi_p^-$ is a sum of three susceptibilities $\chi_{c1,a1}^-$, $\chi_{c2,a2}^-$, $\chi_{c3,a3}^-$, given by Eqs. (6a)-(6c), while $\chi_p^+$ is a sum of three different susceptibilities $\chi_{c3,a1}^+$, $\chi_{c4,a2}^+$, $\chi_{c5,a3}^+$, given by Eqs. (6d)-(6f). Each susceptibility can be seen as the product of three quantities: the square of dipole moment $\mu_{ci,aj}^2$, the frequency detuning factor $F_i$, and the ground-state population $\rho_{ai,ai}$, as given in Eqs.(6). Each dipole moment for a given transition between two Zeeman sub-levels is dependent on a specific atomic C-G coefficient. Each frequency detuning factor $F_i$ depends on the specific three-level EIT sub-system. F1 to F5 (with EIT) are significantly different from F6 (without EIT) near EIT resonance since they are greatly modified by EIT resonance. The three ground-state populations can be calculated with the full density-matrix equations, which depend sensitively on several parameters. The difference between susceptibilities $\chi_p^+$ and $\chi_p^-$ for the two circularly-polarized probe components results in the polarization rotation of the linearly-polarized probe beam. Asymmetry in the number of EIT sub-systems for the two circularly-polarized probe components and differences in the transition strengths (due to different C-G coefficients) both contribute to the degree of polarization rotation. The interplay between these two mechanisms is a complex issue and will be addressed later in Section V.

The maximum achievable polarization rotation angles depend sensitively on the number of the Zeeman sublevels involved (and therefore the numbers of EIT sub-systems for the left- and right-circularly-polarization probe beams) and their transition strengths. Here, we experimentally and theoretically studied the polarization rotation of an optical field in D1 line in rubidium atoms. D2 line in rubidium atoms will not increase the achievable polarization rotation angles due to similar C-G coefficients and furthermore other closely located energy levels ($5P_{3/2}$, F'=1, and F'=3) to F'=2 make the experimental investigation and theoretical analyses much more complicated.

## III. Experimental setup for detecting polarization rotation

Figure 4 depicts the experimental setup. DL1 (probe beam) and DL2 (coupling beam) are both frequency stabilized diode lasers with grating feedback. The probe beam is linearly polarized in the s direction and the coupling beam is left-circularly-polarized by using a polarization beam splitter (PBS2) and a quarter-wave plate. The atomic cell is 5 cm long with magnetic shielding and is temperature stabilized to achieve desired atomic density. The coupling and probe beams co-propagate through the Rb vapour cell. The coupling beam is aligned at a small angle (about $2^o$ ) from the probe beam and they overlap well inside the Rb cell. The power of the probe beam entering the Rb cell is 150 μW, which gives a Rabi frequency of $\Omega_p=2\pi\times10$ MHz at the center of the Rb cell. The probe transmission is split into two parts by a 50/50 beam splitter (BS) whose reflectivity is balanced for s and p linearly-polarized laser beams. A polarized beam splitter (PBS3) splits the two polarization components of the transmitted probe beam from BS into detectors D1 and D2, and another polarized beam splitter (PBS4) splits the reflected probe beam from BS (after it passes through a half-wave plate set with the polarization axis $22.5^o$ from the input probe polarization direction.) into D3 and D4. In the absence of the atomic cell, the polarization direction of the probe beam does not have any rotation. The transmitted probe beam from BS is then fully reflected into D2 by PBS3 and D1 will not detect any signal, so the light signal detected by D1 and D2 are zero and $I_0$ (the intensity of the transmitted probe beam from BS), respectively. The reflected probe beam from BS splits equally into D3 and D4 (since the polarization of the reflected probe beam is rotated by an angle of $45^o$ when it pass through the half-wave plate), so both signals detected by D3 and D4 are $I_0/2$. When the polarization of the probe beam rotates an angle ϕ in Rb cell, the detectors D1, D2, D3 and D4 will detect the signals which are directly related to the rotation angle. If the absorption is not considered, the signals detected by D1, D2, D3 and D4 will be $I_0\sin^2\phi$, $I_0\cos^2\phi$, $I_0\sin^2 (45^o-\phi)$ and $I_0\cos^2 (45^o-\phi)$, respectively. Under the condition of considering the absorption of Rb atoms, the expressions for the signals detected by D1, D2, D3 and D4 are complex, and will be derived next.

The probe beam is assumed to propagate along z axis and its initial polarization direction is in s plane at $0^o$ angle with respect to the x axis. The input field before entering the atomic cell can be written as:

$$E_{in} = \begin{bmatrix} E_{in-x} \\ E_{in-y} \end{bmatrix} = E_0 \begin{bmatrix} \cos 0 \\ \sin 0 \end{bmatrix} = E_0 \begin{bmatrix} 1 \\ 0 \end{bmatrix}, \tag{11}$$

which can be expressed in the circular polarization base vectors as

$$E_{in} = E_0 \begin{bmatrix} 1 \\ 0 \end{bmatrix} = E_0 \left\{ \frac{1}{2} \begin{bmatrix} 1 \\ i \end{bmatrix} + \frac{1}{2} \begin{bmatrix} 1 \\ -i \end{bmatrix} \right\}. \tag{12}$$

The electric field of the probe beam after the cell of length d is changed to [21]

$$E_{out} = E_0 \left\{ \frac{1}{2} \begin{bmatrix} 1 \\ i \end{bmatrix} e^{-i\frac{\omega_p}{c} n_+ d} e^{-\alpha_+ d/2} + \frac{1}{2} \begin{bmatrix} 1 \\ -i \end{bmatrix} e^{-i\frac{\omega_p}{c} n_- d} e^{-\alpha_- d/2} \right\}, \tag{13}$$

where $n_\pm$ are the refractive indices of the Rb atomic medium for the left- and right-circularly-polarized components; $\alpha_\pm$ are the corresponding absorption coefficients.

The transmitted probe beam from BS splits into two parts by PBS3 due to the polarization rotation of the probe beam in the Rb cell, which are detected by detectors D1 and D2, respectively. The intensities detected by D1 and D2 can be expressed as

$$I_{D1} = \frac{1}{2} I_0 \left( \frac{1}{4} e^{-\alpha_+ d} + \frac{1}{4} e^{-\alpha_- d} - \frac{e^{-(\alpha_+ + \alpha_-)d/2}}{2} \cos\left[ \frac{2\pi}{\lambda} (n_+ - n_-) d \right] \right),$$

$$I_{D2} = \frac{1}{2} I_0 \left( \frac{1}{4} e^{-\alpha_+ d} + \frac{1}{4} e^{-\alpha_- d} + \frac{e^{-(\alpha_+ + \alpha_-)d/2}}{2} \cos\left[ \frac{2\pi}{\lambda} (n_+ - n_-) d \right] \right). \tag{14}$$

$I_0$ is the intensity of the input probe beam. The intensity difference between D1 and D2 is given by

$$I_{D1} - I_{D2} = -\frac{1}{2} I_0 e^{-(\alpha_+ + \alpha_-)d/2} \cos[2\phi]. \tag{15}$$

After the half-wave plate, the electric field of the reflected probe beam from BS can be expressed as

$$E'_{out} = \begin{pmatrix} -\frac{\sqrt{2}}{2} & \frac{\sqrt{2}}{2} \\ \frac{\sqrt{2}}{2} & \frac{\sqrt{2}}{2} \end{pmatrix} \begin{pmatrix} \frac{1}{\sqrt{2}} \mathrm{E}_{\text{out-x}} \\ \frac{1}{\sqrt{2}} \mathrm{E}_{\text{out-y}} \end{pmatrix}, \tag{16}$$

which is split into detectors D3 and D4 by PBS4. When the indices of refraction $n_-$

and $n_+$ are different, the intensities detected by D3 and D4 are given by

$$I_{D3} = \frac{1}{2} I_0 \left( \frac{1}{4} e^{-\alpha_+ d} + \frac{1}{4} e^{-\alpha_- d} - \frac{e^{-(\alpha_+ + \alpha_-) d/2}}{2} \sin\left[ \frac{2\pi}{\lambda} (n_+ - n_-) d \right] \right) ,$$

$$I_{D4} = \frac{1}{2} I_0 \left( \frac{1}{4} e^{-\alpha_+ d} + \frac{1}{4} e^{-\alpha_- d} + \frac{e^{-(\alpha_+ + \alpha_-) d/2}}{2} \sin\left[ \frac{2\pi}{\lambda} (n_+ - n_-) d \right] \right) , \qquad (17)$$

which give the intensity difference between D3 and D4 as

$$I_{D3} - I_{D4} = -\frac{1}{2} I_0 e^{-(\alpha_+ + \alpha_-) d/2} \sin[2\phi] . \qquad (18)$$

So, after the probe beam propagates through the atomic cell which contains atomic medium, its polarization rotation angle $\phi$ is given by

$$\phi = \frac{1}{2} arctg\left[ \frac{I_{D3} - I_{D4}}{I_{D1} - I_{D2}} \right] . \qquad (19)$$

Comparing to the conventional measurement scheme [21], one can see that the expression of the polarization rotation angle $\phi$ in our measurement scheme is independent of absorption of the atomic medium since we used a 50% beam-splitter to split the polarization rotated probe beam into two polarization measurement systems. This scheme allows us to eliminate the effect of absorption on polarization rotation angle $\phi$, which is a great advantage since the variation of absorption near the edges of an EIT window is quite large.

## IV. Results and discussion

In the experiment, the coupling beam frequency was locked to the atomic transition from level $5S_{1/2}$, F=2 to level $5P_{1/2}$, F'=2, while the frequency of the probe beam was scanned around the transition from level $5S_{1/2}$, F=1 to level $5P_{1/2}$, F'=2. We first checked the asymmetry in the number of EIT sub-systems for the two circularly-polarized probe components by adding a magnetic field (~10 G) in the z-direction of the atomic cell (parallel to the laser beam propagation direction). When the probe beam had only $\sigma^-$ component (by using a quarter-wave plate, not shown, in the probe beam before entering the cell), three EIT peaks were observed in the probe transmission (with $\sigma^-$ coupling beam) corresponding to the three simple EIT sub-systems, as shown in Fig. 5(a). As the magnetic field was turned off, a single degenerate EIT peak, as shown in Fig. 5(b), was recorded. However, if the probe beam had only $\sigma^+$ component (with the $\sigma^-$ coupling beam and a magnetic field of ~10 G), only two EIT peaks were observed, as shown in Fig. 5(c). Again, by turning off the magnetic field, one degenerate EIT peak was measured, as shown in Fig. 5(d). The relative heights of the EIT peaks in Figs. (a) and (c) are determined by the differences in the transition strengths connecting different Zeeman sublevels. It is clear from Figs. 5(b) and (d) that the total EIT widths and heights for these two circularly-polarized probe components (without the magnetic field) are quite different due to the asymmetry in the number of degenerate EIT sub-systems and contributions from differences in transition strengths (due to different C-G coefficients). This indicates that a linearly-polarized probe beam will experience strong birefringence, especially at the edges of the EIT windows, as given by Eq. (10).

Next, we measured the polarization rotation angle $\phi$ of the probe beam. Without the coupling beam, the s-polarized probe beam will be totally reflected by PBS3 and no light will be detected by D1, at the same time, the powers of the probe beam entering detectors D3 and D4 are balanced. This indicates that the polarization of the probe beam is not rotated. As the left-circularly-polarized coupling beam is turned on, the polarization of the probe beam is rotated by an angle $\phi$ which can be determined by the four detectors D1, D2, D3 and D4, as given in Eq. (19). The part of the transmitted probe beam from BS will pass PBS3 and be detected by D1, as shown in Fig. 6(a). Figure 6(b) is the intensity detected by D2. Also, the powers of the reflected probe beam from BS entering the detectors D3 and D4 become not balanced. As shown in Figs. 6(c) and (d), near probe resonant frequency ($\Delta\omega_p \approx 0$), a sharp change appears in Fig. 6(c) (detected by D3) and a reversed profile appears in Fig. 6(d) (detected by D4), respectively. With the experimental data detected by D1, D2, D3 and D4, we can calculate the degree of polarization rotation from Eq. (19) as a function of probe frequency detuning.

We have also studied the polarization rotation as a function of coupling beam power and the results are plotted in Fig. 7, in which Figs. 7(a1), (a2), (a3) are for coupling powers of $P_c$= 6, 10, 15 mW, respectively. One can see that there are two dispersion-like peaks (one up and one down) near the two-photon resonance condition. The rotation angles at both peaks increase with the coupling beam power, but at difference rates. Such asymmetry is partly caused by different ac Stark shifts involving the additional energy levels of $5P_{1/2}$, F'=1, which give different coupling beam frequency detunings for different transitions and make the centers of transmission and dispersion profiles of the Λ-type EIT sub-systems shift by different values. Similar to the phenomenon of linear magneto-optical rotation [3], the asymmetric ac Stark shifts also cause optical birefringence for the probe beam. Figures 7(b1), (b2), (b3) plot the corresponding theoretical calculations of the polarization rotation angle $\phi$ as a function of the probe beam detuning for $P_c$=6, 10, 15 mW (corresponding to the Rabi frequencies of $\Omega_c=2\pi\times63$ MHz, $2\pi\times82$ MHz, $2\pi\times100$ MHz, respectively) with the Doppler effect included. The shapes and peak values of Figs. 7(b1), (b2), (b3) are in good agreements with the experimentally measured results as shown on Figs. 7(a1), (a2), (a3), respectively.

Figure 8 presents the maximal polarization rotation angles of the two peaks at different probe beam frequency detunings as a function of the coupling power [16]. As one can see that the agreements between the theoretically calculated results (curves a and d) and experimentally measured data (curves b and c) are quite good. The remaining discrepancies come from the imprecise calculation for the ground-state populations, i.e. the ground-state populations were calculated by solving Eq.3 without considering the Doppler broadening. Also, the spatial variations of the probe and coupling laser beams, the propagating losses of the laser beams in the atomic cell, and high-order contributions of the probe beam power were not considered in the theoretical calculations.

Figure 9 plots the experimentally measured and theoretically calculated polarization rotation angles of the probe beam at different temperatures of the Rb vapor cell. The coupling power is about 15 mW (corresponding to the Rabi frequency of $\Omega_c=2\pi\times100$ MHz). Quite large polarization rotation angle (~$45^o$) has been realized, as shown in Fig. 9(b3), by using a relatively low coupling beam power (15 mW), which gives a significant advantage over previously demonstrated schemes [8-11]. The theoretically calculated degrees of polarization rotation match well with the measured data, especially for the left peak. These results show that large polarization rotation angle can be achieved at higher atomic density without suffering too much absorption due to the use of EIT effect in this new scheme, which can be significant in potential applications.

## V. Contributions to the polarization rotation

One important question is what are the main mechanisms causing such polarization rotation of a linearly-polarized probe beam in this multi-Zeeman-level atomic system. As we have mentioned earlier, two main factors contribute to this polarization rotation. One is the asymmetry in the number of EIT sub-systems for the left- and right-circularly-polarized probe components due to the use of a left-circularly-polarized coupling beam as depicted in Fig. 1. Another main factor is the differences in transition strengths among different Zeeman levels due to the differences in the C-G coefficients [18]. To address the question of which of these mechanisms is the dominant factor in causing the polarization rotation, we consider few altered atomic systems in the following.

First, we consider the same atomic system as the one in Fig. 1, but with a different coupling beam. We let the coupling beam to be linearly polarized and to propagate in the direction perpendicular to the weak mangetic field (could be easily done in cold atoms), which drives the transitions from |bi> to |ci> (solid lines), as shown in Fig. 10. The transition strength from |b3> to |c3> is zero. The probe beam is kept to be same as before, traveling in the direction of the weak magnetic field, with two circularly-polarized components (dashed and dotted lines in Fig. (10)). This system forms four three-level Λ-type EIT sub-systems — two for the left-circularly-polarized probe component and two for the right-circularly-polarized probe component. Since the numbers of EIT subsystems are the same and the transition strengths are also symmetric (although different for different transitions) for the two circularly-polarized probe components, the calculated polarization rotation angle from equations in Section II is zero for any probe detunings ( $\chi_p^+$ and $\chi_p^-$ are the same in Eq. (10)). Therefore, we can conclude that in the total symmetric systems, no polarization rotation can occur, even the C-G coefficients are different. We also experimentally checked the polarization rotation signal by replacing the circularly-polarized coupling beam with a linearly-polarized coupling beam. When the input coupling beam power was 12 mW, temperature of the cell was about 55$^o$C, and the ambient magnetic field in the cell was reduced about 25 mG by magnetic shielding, we observed that the maximum polarization rotation angle for the probe beam is only about 1$^o$. Such small polarization rotation angle probably comes from the residual ambience magnetic field in the atomic cell.

The second considered system is also in the D1 line of $^{87}$Rb atom. By replacing the states of $5P_{1/2}$, F'=2 with $5P_{1/2}$, F'=1, as shown in Fig. 11, and keeping the same linearly-polarized probe beam and left-circularly-polarized coupling beam as in the current experiment, the system is now symmetric in the number of EIT subsystems for the left- and right-circularly-polarized probe components. However, the differences in the C-G coefficients, as shown in Fig. 11 [18], make the transition strengths asymmetric for the two circularly-polarized probe components, which is caused by the left-circularly-polarized coupling beam. By solving the relevant density-matrix equations at steady state, as done in Section II, we can calculate the polarization rotation angle for the probe beam, which is very small as given in Fig. 12 for comparable parameters used in calculating Fig. 2. It only gives a small polarization rotation angle of about $2.5^{o}$ near EIT resonance. Our experimental observation also confirmed such prediction. From studying this system, we can conclude that the asymmetry in the number of EIT subsystems for the two circularly-polarized probe components is the dominant mechanism to cause the polarization rotation of the linearly-polarized probe beam, and the asymmetry in different C-G coefficients alone can only give a minor contribution to the polarization rotation by creating asymmetries in individual EIT subsystems.

The last system to consider is a situation with the same atomic levels and laser beams as in Fig. 1, but to let all the transition strengths to be the same by taking an average value for all transitions (e.g. "to have the same C-G coefficients"). Of course, such atomic system does not exist in nature and can only be considered as a model system. We can calculate the polarization rotation angle for such model system with the set of density-matrix equations as given in Section II. We found that the calculated polarization rotation angle depends very sensitively on the average values chosen for C-G coefficients for different transitions (from |ai> to |ci> for the left-circularly-polarized probe component, from |ai> to |ci+2> for the right-circularly-polarized probe component, and from |bi+1> to |ci> for the left-circularly-polarized coupling beam, respectively), as well as on the steady-state ground-state population distribution. Since it is hard to justify what should be the "correct" values to use as the average C-G coefficients in the calculation, we will not give quantitative results here in this situation. Through various calculations done for different choices of "average C-G coefficients", we can conclude that the differences in the C-G coefficients in different transitions contribute substantially to the polarization rotation when combined with the asymmetry in the number of EIT subsystems for the two circularly-polarized probe components in the current system.

The above studies indicate that the asymmetry in the number of EIT subsystems for the two circularly-polarized probe components is essential in causing the polarization rotation of the probe beam. With such asymmetry in the number of EIT subsystems, the differences in the C-G coefficients can have substantial contribution to the degree of polarization rotation. It is difficult to simply consider these two mechanisms as separate contributions in this case, since they are entangled to produce such large polarization rotation with a relatively low coupling beam power. As one can see that the atomic energy levels and properly chosen laser beams used in the current scheme, as shown in Fig. 1, is quite unique in generating such asymmetry for the two circularly-polarized probe components, therefore creates a large birefringence which can be controlled by the coupling laser power (Fig. 8).

## VI. Conclusion

We experimentally and theoretically studied the phenomenon of polarization rotation of a linearly-polarized optical field controlled by another laser field in the multi-Zeeman-sublevel atomic system. The experimentally measured polarization rotation angles were compared quantitatively with the theoretically calculated results and good agreements were obtained, which helped us to understand the underlying mechanisms for obtaining the large polarization rotation with a low controlling laser power. The large polarization rotation is caused by the asymmetry in the number of EIT subsystems seen by the two circularly-polarized probe components due to the use of the left-circularly-polarized coupling (controlling) beam. The differences in the transition strengths (different C-G coefficients), when combined with such asymmetry in EIT subsystems, also contribute to the degree of polarization rotation in this unique system. The achievable large birefringence with a relatively low controlling optical power in this system makes it potentially very useful as dynamic polarization elements (such as wave-plates) in atomic assembles. By exploiting the EIT property in this scheme, the problem of strong circular dichroism is avoided, which gives a great advantage over previously demonstrated schemes for controlling polarization rotation in atomic samples. Due to the sharp change in the polarization rotation angle as a function of the probe detuning and low absorption due to EIT, a controlled all-optical switch (or logic gate) can be constructed in this system with small frequency detuning.

We acknowledge funding supports by the CNSF (# 60325414, 60578059, 60238010 and RGC60518001) in China and PCSIRT, the CFKSTIP (Ministry of Education of China, # 705010). *Corresponding author H. Wang's e-mail address is wanghai@sxu.edu.cn.

## References:

[1] D. Budker et al., Rev. Mod. Phys., **74**, 1153 (2002); and references therein.

[2] I. Novikova, A.B. Matsko, and G. R. Welch, Opt. Lett., **26**, 1016 (2001).

[3] C. Cohen-Tannoudji and J. Dupont-Roc, Phys. Rev. A **5**, 968 (1972).

[4] P.F. Liao and G.C. Bjorklund, Phys. Rev. Lett., **36**, 584 (1976).

[5] S.E. Harris, Physics Today, **50**, 36 (1997).

[6] J. Gea-Banacloche, Y. Li, S. Jin, and Min Xiao, Phys. Rev. A **51**, 576 (1995).

[7] Min Xiao, Y. Li, S. Jin, and J. Gea-Banacloche, Phys. Rev Lett., **74**, 666 (1995).

[8] F.S. Pavone et al., Opt. Lett., **22**, 736 (1997).

[9] S. Wielandy and A.L. Gaeta, Phys. Rev. Lett., **81**, 3359 (1998).

[10] T.H. Yoon, C.Y. Park, and S.J. Park, Phys. Rev. A **70**, 061803(R) (2004).

[11] S.J. Park, C.Y. Park, and T.H. Yoon, Phys. Rev. A **71**, 063819 (2005).

[12] D. Cho, J. M. Choi, J. M. Kim and Q. H. Park, Phys. Rev. A **72**, 023821 (2005).

[13] A.K. Patnaik and G.S. Agarwal, Opt. Commun., **179**, 97 (2000).

[14] G.S. Agarwal and S. Dasgupta, Phys. Rev. A **67**, 023814 (2003).

[15] V.A. Sautenkov et al., Phys. Rev. Lett., **94**, 233601 (2005).

[16] B. Wang, S. Li, J. Ma, H. Wang, K.C. Peng, and Min Xiao, Phys. Rev. A, Rapid Communications, **73**, 051801(R) (2006).

[17] Y. Li and Min Xiao, Phys. Rev. A **51**, 2703 (1995).

[18] For detail in transition probabilities of D1 line in $^{87}$Rb, see http://steck.us/alkalidata.

[19] Hong Yuan Ling, Y.Q. Li and Min Xiao, Phys. Rev. A, **53**, R1014 (1996).

[20] M. Weidemuller and C. Zimmermann, *Interactions in Ultracold Gases* WILEY-VCH, 2003).

[21] C P Pearman et al., J. Phys. B: At. Mol. Opt. Phys., **35**, 5141 (2002).

## Figure Captions:

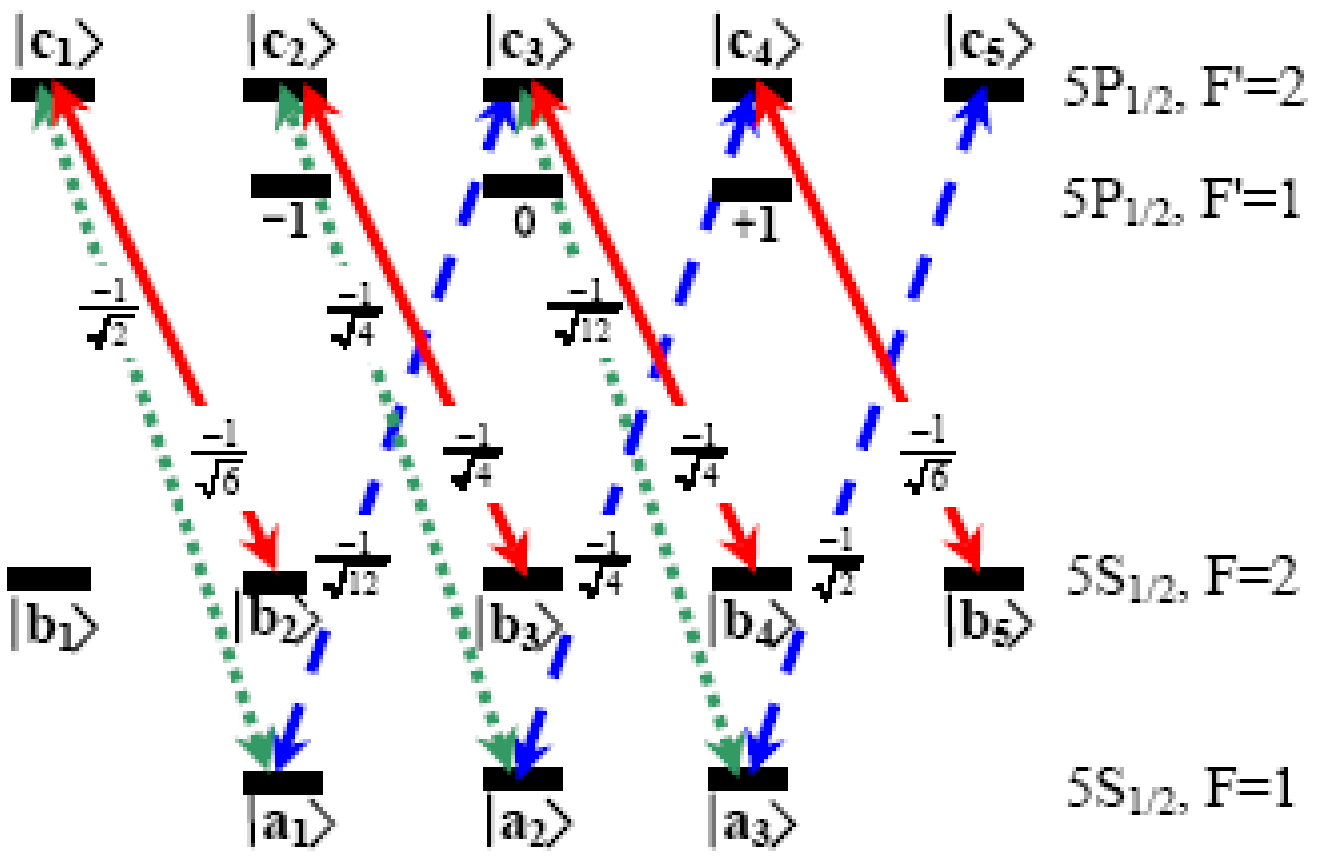

FIG. 1. Relevant energy level diagram of the D1 line in $^{87}$Rb atom. Solid lines: transitions for the left-circularly-polarized coupling beam; dotted lines: transitions for the left-circularly-polarized probe beam; dashed lines: transitions for the right-circularly-polarized probe beam.

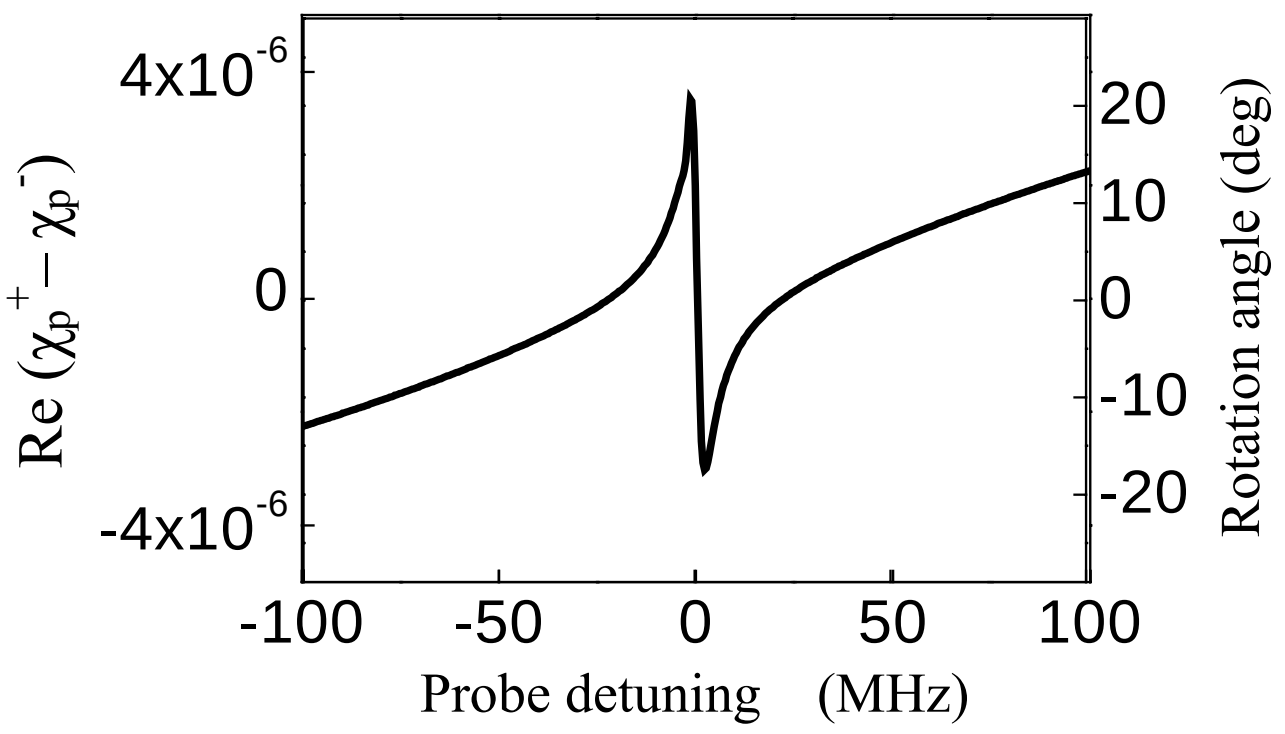

FIG. 2. The difference between Re($\chi_p^+$) and Re($\chi_p^-$) (right vertical axis is the polarization rotation angle). The parameters used here are $\Omega_c$=2π×80 MHz, $\Omega_p$=2π×10 MHz, $\gamma_{ac}$=2π×3.5 MHz and $\gamma_{ab}$=2π×1.1 MHz, atomic density of N= 1.8×$10^{11}$/cm$^3$ .

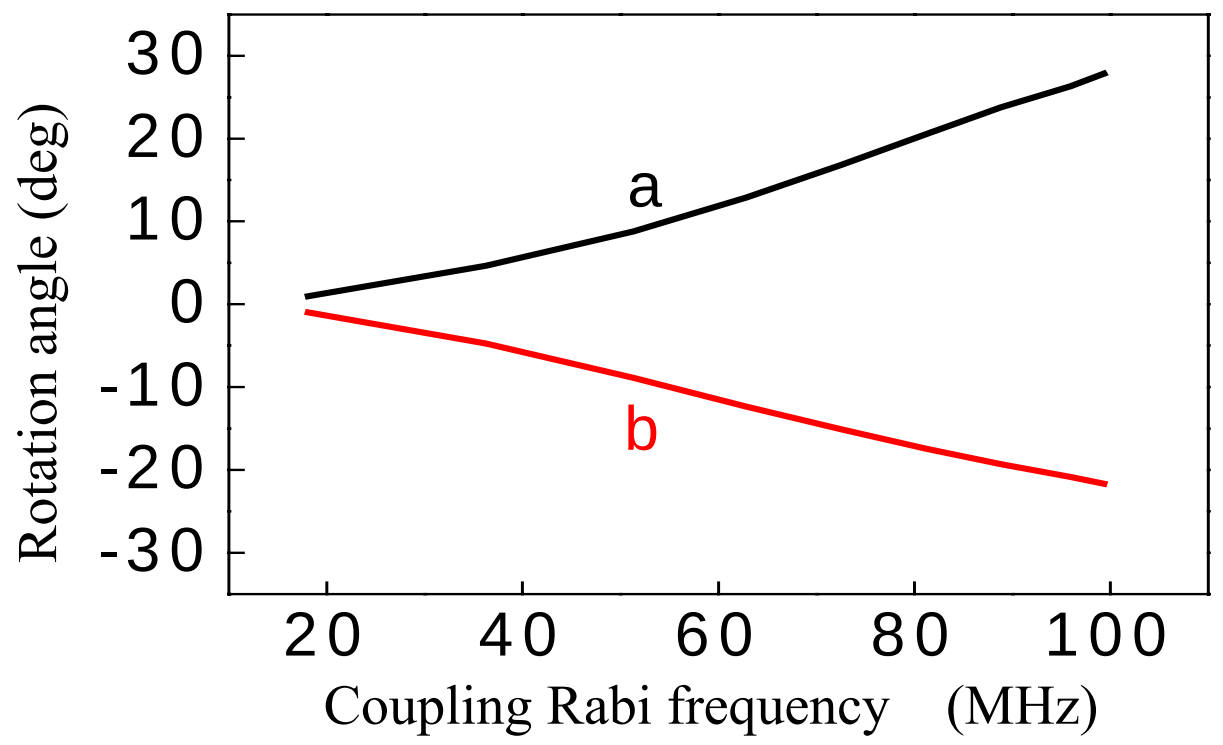

FIG. 3. Theoretically calculated results of the polarization rotation angle as a function of coupling Rabi frequency. Curves (a) and (b) are results for the left and right dispersion peaks, as shown in Fig.2, respectively. The parameters are $\Omega_p=2\pi\times10$ MHz, $\gamma_{ab}=2\pi\times1.1$ MHz , $\gamma_{ac}=2\pi\times3.5$ MHz, and atomic density of $N=1.8\times10^{11}/cm^3$.

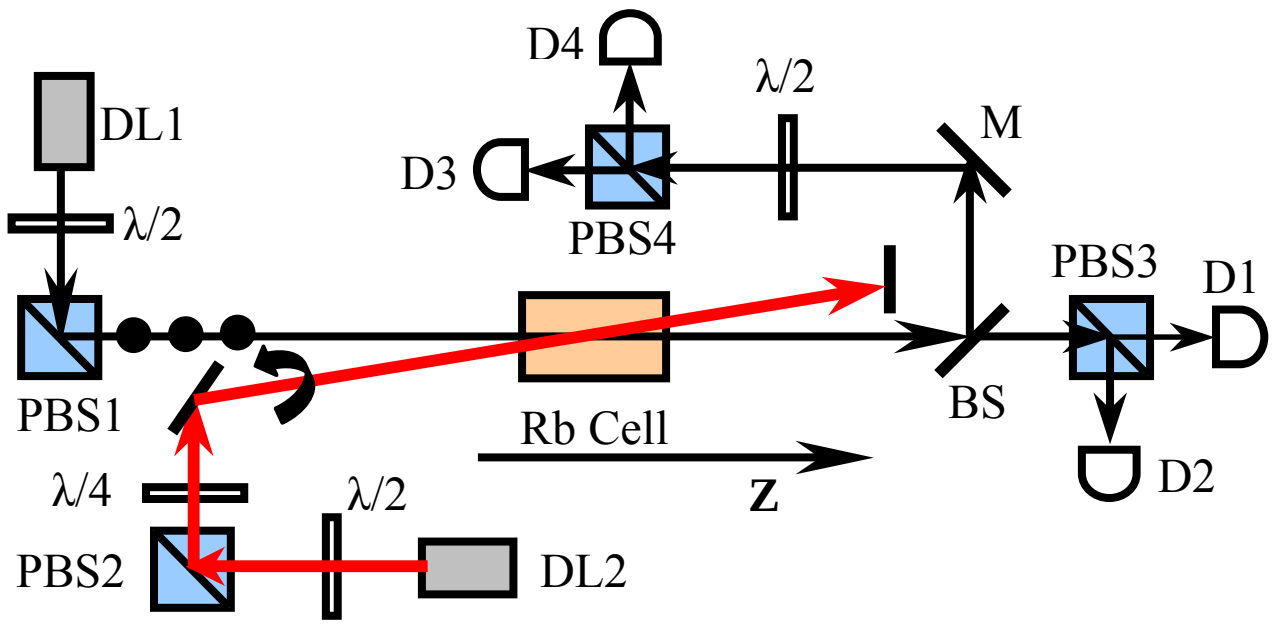

FIG. 4. Experimental set-up. DL1 and DL2: diode lasers; PBS1-PBS4: polarization cube beam splitters; λ/2 and λ/4: half-wave and quarter-wave plates; D1-D4: photodetectors.

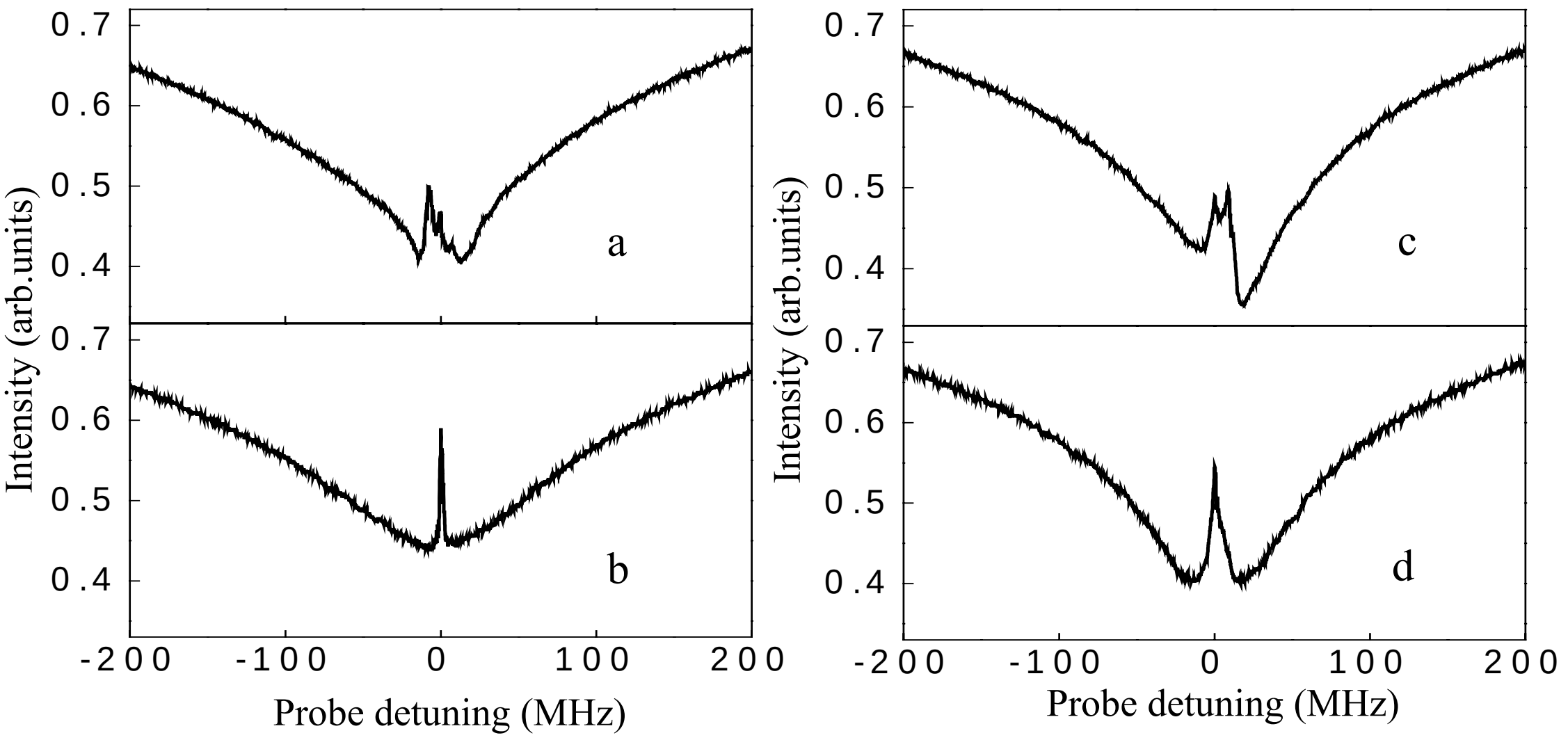

FIG. 5. EIT curves for (a) the left-circularly-polarized probe beam with magnetic field; (b) the left-circularly-polarized probe beam without magnetic field; (c) the right-circularly-polarized probe beam with magnetic field; (d) the right-circularly-polarized probe beam without magnetic field. The coupling beam is left-circularly-polarized. The probe beam power is 75 μW and the power of the coupling beam is 6 mW. The temperature of cell is 55°C.

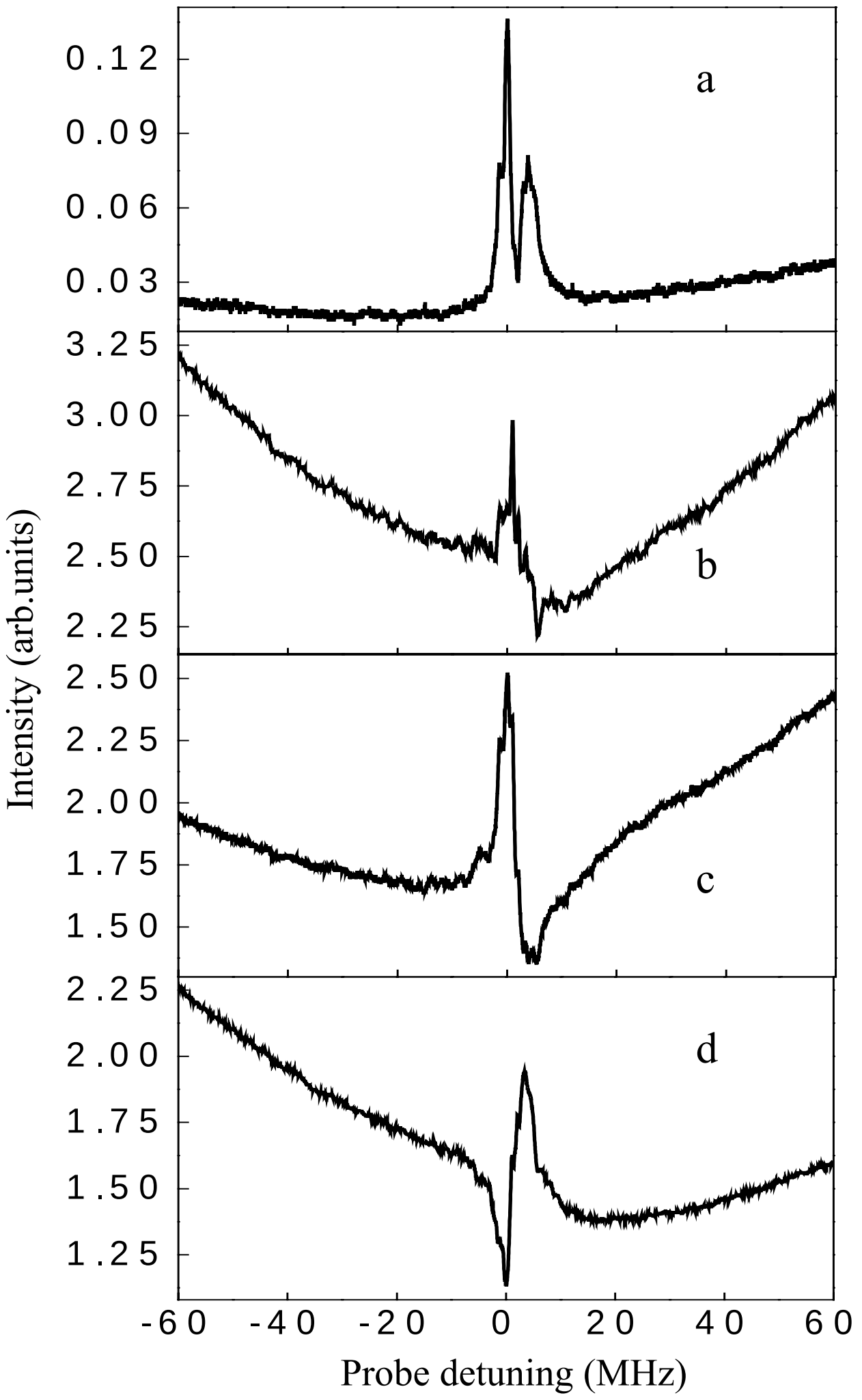

FIG. 6. Experimentally detected intensity signals as a function of probe detuning by detectors (a) D1; (b) D2; (c) D3; (d) D4, respectively. The temperature of the atomic cell T=50$^{o}$C, the power of the coupling beam $P_C$=15 mW, the power of the probe beam $P_p$=150 μW, and the coupling detuning $\Delta_c$=0.

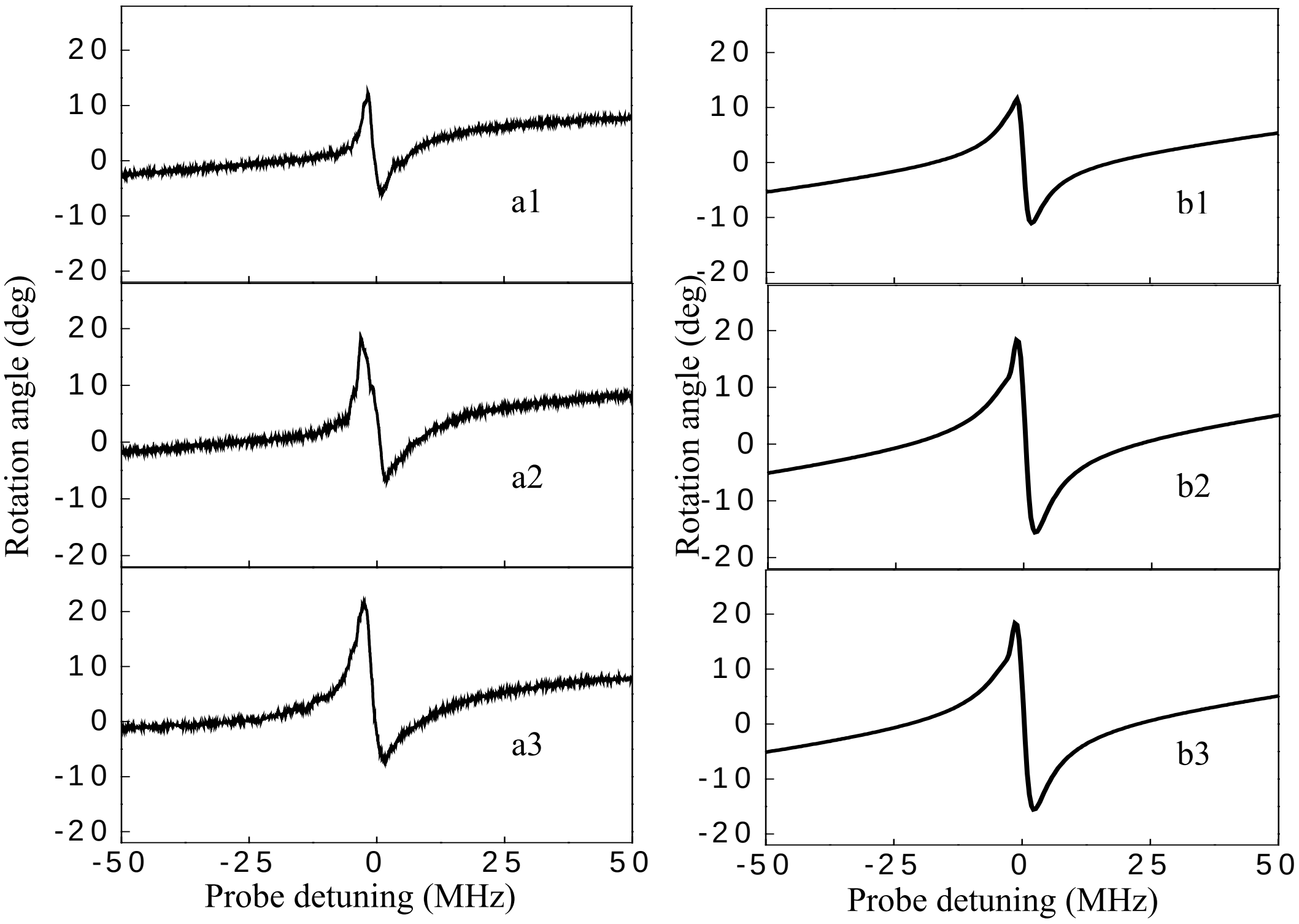

FIG. 7. The experimental and theoretical results for the rotation angle as a function of probe detuning at temperature T=55$^{o}$C. (a1), (a2), (a3) are the measured results for the coupling power $P_c$= 6, 10, 15 mW, and (b1), (b2), (b3) are the calculated results from Eqs. (6)-(10) for the same coupling powers, respectively.

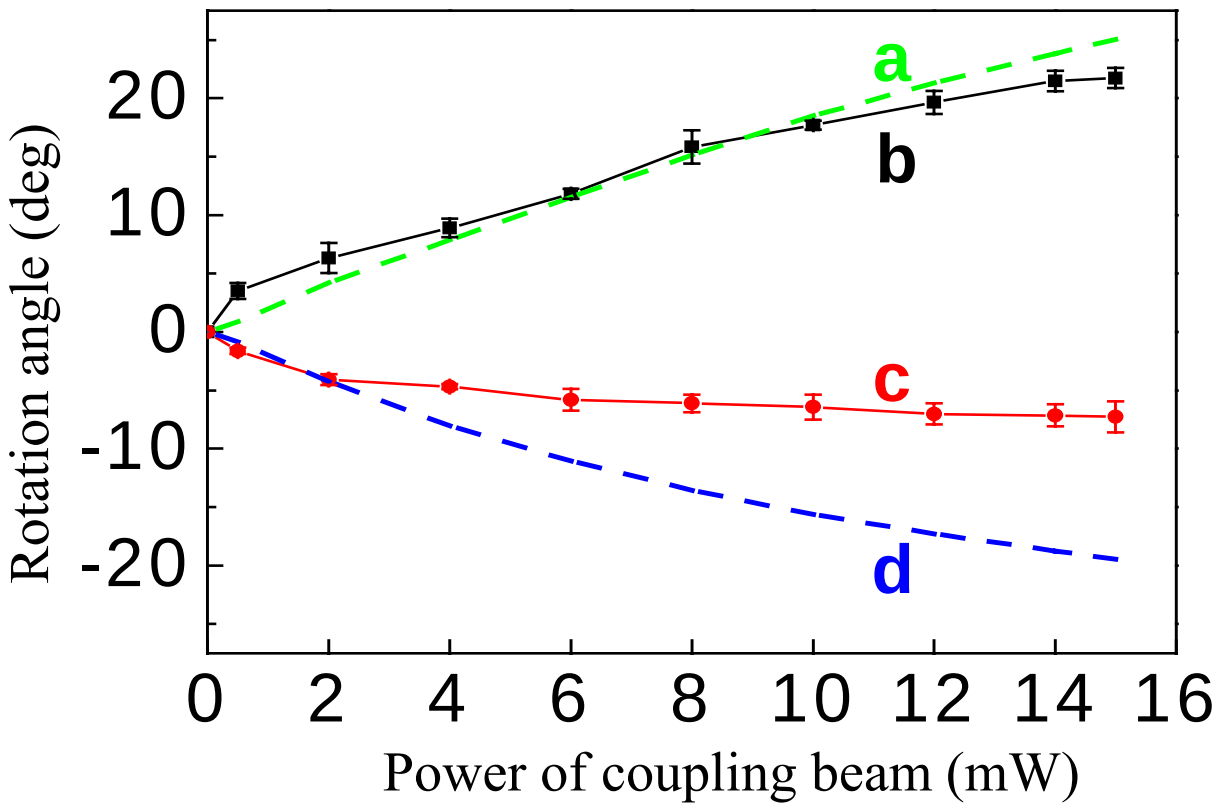

FIG. 8. Degree of polarization rotation as a function of the coupling beam power. Curves (b) and (c) are experimental results for the left and right dispersion-like peaks, respectively. The temperature of the atomic cell T=55°C; Curves (a) and (d) are theoretical results for the left and right dispersion-like peaks, respectively, with parameters $\gamma_{ab}=2\pi\times1.1$ MHz , $\gamma_{ac}=2\pi\times3.5$ MHz , and $N=1.62\times10^{11}/cm^3$.

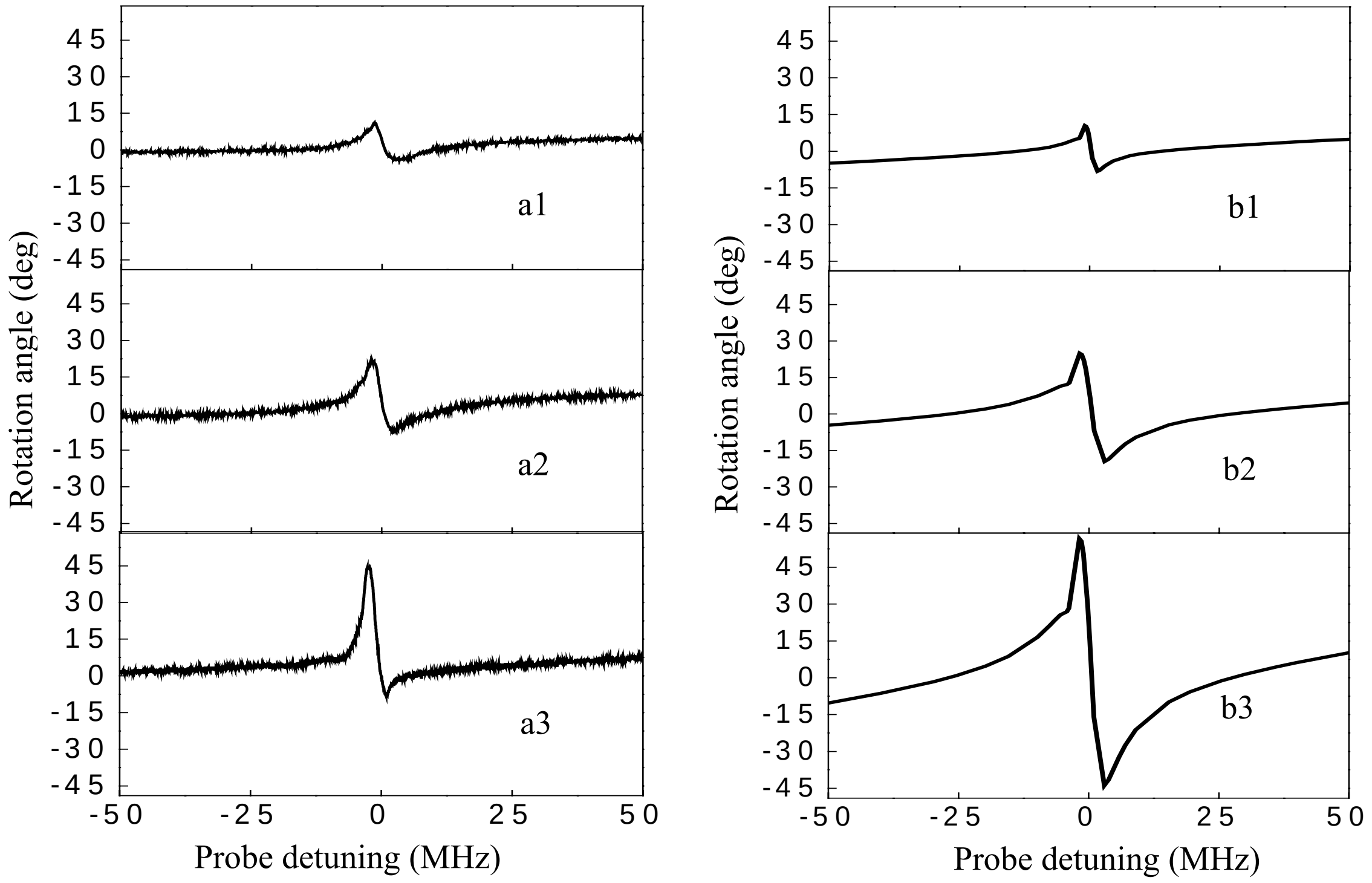

FIG. 9. Angle of polarization rotation for temperatures of Rb vapor cell at T=45°, 55°, 65°, respectively. Figures (a1), (a2) and (a3) are experimental results for coupling power of 15 mW ($\Omega_c=2\pi\times100$ MHz); Figures (b1), (b2) and (b3) correspond to theoretical results with parameters $\gamma_{ab}=2\pi\times1.1$ MHz, $\gamma_{ac}=2\pi\times3.5$ MHz, and $\Omega_c$ $=2\pi\times100$ MHz.

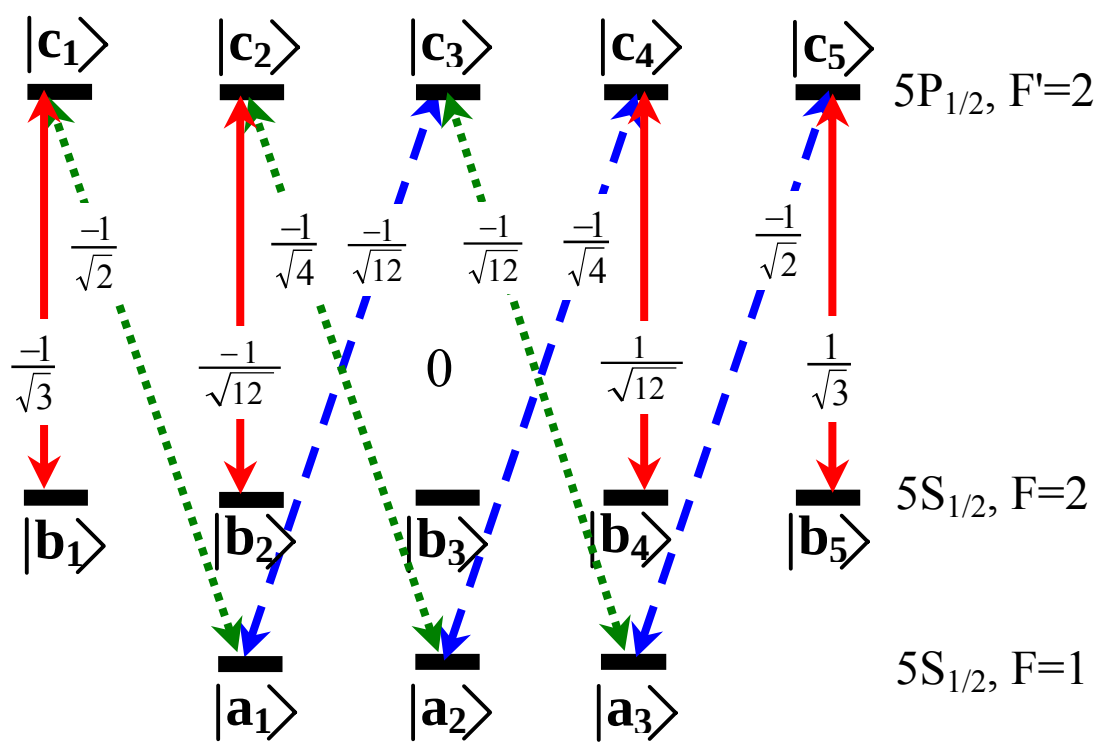

Fig. 10. Relevant energy level diagram of D1 line in [87]Rb atom. Solid lines: transitions for the coupling beam; dotted lines: transitions for the left-circularly-polarized probe beam; dashed lines: transitions for the right-circularly-polarized probe beam.

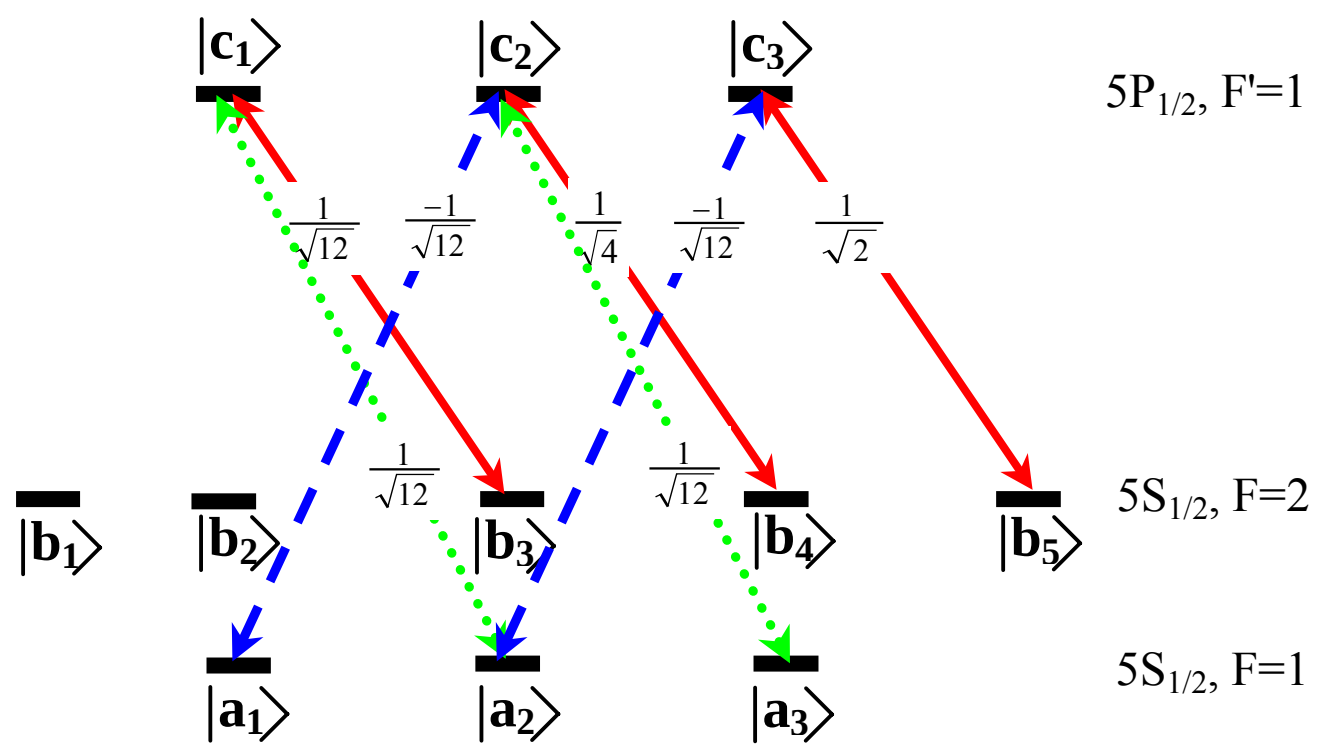

Fig. 11. Relevant energy diagram of D1 line in [87]Rb atom. Solid lines: transitions for the left-circularly-polarized coupling beam; dotted lines: transitions for the left-circularly-polarized probe beam; dashed lines: transitions for the right-circularly-polarized probe beam.

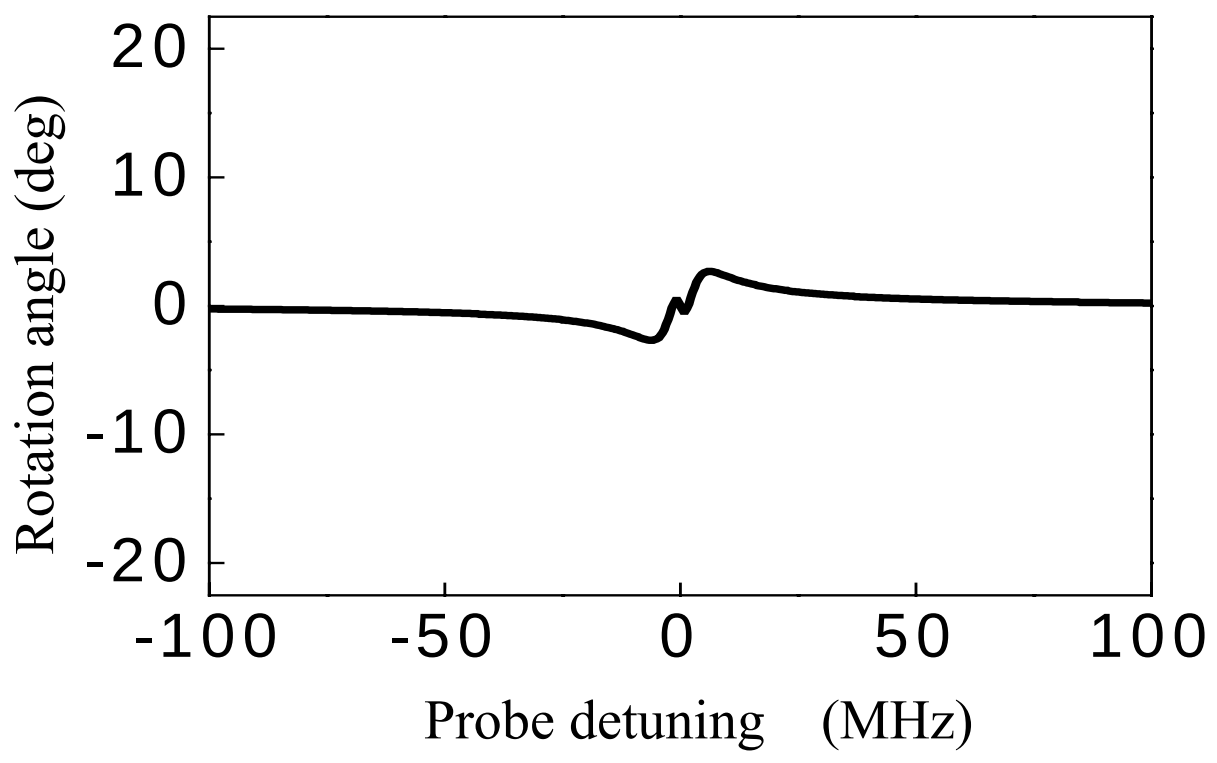

Fig. 12. Calculated polarization rotation angle as a function of probe detuning for the system in Fig.11. The parameters used in the calculation are the same as in Fig.2 (without considering the AC stark effect).